# A Comparative Dosimetric Study of Proton and Photon Therapy in Stereotactic Arrhythmia Radioablation for Ventricular Tachycardia

Keyur D. Shah, PhD[1#], Chih-Wei Chang, PhD[1#], Pretesh Patel, MD[1], Sibo Tian, MD[1], Yuan Shao, PhD[2], Kristin A Higgins, MD[3], Yinan Wang MS[1], Justin Roper, PhD[1], Jun Zhou, PhD[1], Zhen Tian, PhD[4] and Xiaofeng Yang, PhD[1*]

[1]Department of Radiation Oncology and Winship Cancer Institute, Emory University, Atlanta, GA

[2]School of Public Health, University of Illinois, Chicago, IL

[3]Department of Radiation Oncology, City of Hope Cancer Center Atlanta, Atlanta, GA

[4]Department of Radiation & Cellular Oncology, University of Chicago, Chicago, IL

[#]Co-first author

[*]Corresponding to: xiaofeng.yang@emory.edu

**Abstract**

**Purpose:** Ventricular tachycardia (VT) is a life-threatening arrhythmia commonly treated with catheter ablation; however, some cases remain refractory to conventional treatment. Stereotactic arrhythmia radioablation (STAR) has emerged as a non-invasive option for such patients. While photon-based STAR has shown efficacy, proton therapy offers potential advantages due to its superior dose conformity and sparing of critical organs at risk (OARs), including the heart itself. This study aims to investigate and compare the dosimetry between proton and photon therapy for VT, focusing on target coverage and OAR sparing.

**Methods:** We performed a retrospective study on a cohort of 34 VT patients who received photon STAR. Proton STAR plans were generated using robust optimization in RayStation to deliver the same prescription dose of 25 Gy in a single fraction while minimizing dose to OARs. Dosimetric metrics, including $D_{99}$, $D_{95}$, $D_{mean}$, and $D_{0.03cc}$, were extracted for critical OARs (heart, lungs, esophagus) and ventricular arrhythmogenic substrates (VAS). Shapiro-Wilk tests were used to assess normality, followed by paired t-tests or Wilcoxon signed-rank tests for statistical comparisons between modalities, with Bonferroni correction applied for multiple comparisons.

**Results:** Proton and photon plans achieved comparable target coverage, with VAS $D_{95}$ of 24.1 ± 1.2 Gy vs. 24.7 ± 1.0 Gy (p=0.294). Proton therapy significantly reduced OAR doses, including heart $D_{mean}$ (3.6 ± 1.5 Gy vs. 5.5 ± 2.0 Gy, p<0.001), lungs $D_{mean}$ (1.6 ± 1.5 Gy vs. 2.1 ± 1.4 Gy, p<0.001), and esophagus $D_{mean}$ (0.3 ± 0.6 Gy vs. 1.6 ± 1.3 Gy, p<0.001), while maintaining optimal target coverage.

**Conclusion:** Proton therapy for STAR demonstrates significant dosimetric advantages in sparing the heart and other critical OARs compared to photon therapy for VT, while maintaining equivalent target coverage. These findings highlight the potential of proton therapy to reduce treatment-related toxicity and improve outcomes for VT patients.

1. Introduction

Ventricular tachycardia (VT) is a heart rhythm disorder caused by abnormal electrical signals in the lower chambers of the heart (ventricles) (1, 2). VT commonly occurs in individuals with underlying heart conditions, such as prior myocardial infarction or structural heart disease (e.g., cardiomyopathy) (3). Between 2007 and 2020, ventricular tachycardia (VT) contributed to 7,025 deaths in the United States among patients with underlying cardiovascular disease (CVD) (4). Current treatment options for VT include implantable cardioverter defibrillators (ICDs), antiarrhythmic medications, and catheter ablation, often used in combination to manage the condition. Catheter ablation, introduced over 25 years ago, remains the gold standard for treating drug-refractory VT by targeting arrhythmogenic substrates with localized radiofrequency energy via intracardiac catheters (5, 6). However, catheter ablation is not curative for many patients, particularly those with inaccessible arrhythmogenic tissue or inadequate energy delivery across the myocardial wall. Additionally, the procedure is associated with high rates of complications and even mortality, particularly in patients with advanced heart failure or extensive comorbidities (7, 8).

Recent advances in heart failure therapies have extended survival in patients with severe cardiac dysfunction. However, these patients frequently develop VT or ventricular fibrillation (VF) because of progressive myocardial disease. Treating VT in this population remains challenging due to the complexity and extent of arrhythmogenic substrates, as well as the basal and epicardial locations of VT origins that are challenging to access (9). This has driven interest in exploring alternative, less invasive approaches for VT treatment. Stereotactic body radiation therapy (SBRT) has recently emerged as a novel, non-invasive approach for managing VT. By delivering high-dose radiation to the ventricular arrhythmia substrate (VAS), SBRT offers the potential to achieve arrhythmia modulation without the risks associated with catheter-based interventions (10). Initially developed for treating oncologic targets, SBRT has demonstrated precision and efficacy in delivering ablative doses to various anatomical targets. For VT, this approach, referred to as stereotactic arrhythmia radioablation (STAR), applies SBRT techniques specifically to cardiac arrhythmogenic substrates, offering a promising alternative for patients who have exhausted other therapeutic options. While the mechanisms of action differ from catheter ablation, the ability of radiation to induce fibrosis and alter myocardial cell function provides a foundation for VT management. However, vascular and inflammatory effects induced by high-dose radiation remain poorly understood, necessitating further study (11, 12).

Extending STAR from photons to protons introduces additional considerations. Proton therapy, characterized by the Bragg peak phenomenon, allows for precise dose delivery with minimal exit dose, potentially improving sparing of adjacent critical structures such as the lungs, esophagus, stomach, and ribs. Preclinical studies have explored the use of high-energy particles for arrhythmia ablation, including carbon ions and protons (13–16). While carbon ion therapy offers superior radiobiological effectiveness, its cost and limited availability make protons a more accessible alternative. A first-in-man case report demonstrated the feasibility and safety of proton STAR, achieving a significant reduction in VT burden in a patient with advanced heart failure and refractory VT (17). Despite these promising results, challenges such as range uncertainties and motion management remain significant barriers to its widespread adoption.

This study evaluates the dosimetric trade-offs between proton- and photon-based STAR for VT, focusing on target coverage, dose conformity, and sparing of critical organs at risk. By analyzing clinically derived photon STAR plans alongside retrospectively developed proton plans, this work aims to guide clinical implementation and identify patient-specific factors that could influence modality selection in VT management.

## 2. Methods

### 2.1 Patient and Target Selection

Patients included in this retrospective dosimetric study were treated under compassionate-use approval from the institutional review board. Inclusion criteria required a confirmed diagnosis of refractory VT with documented recurrence despite at least two failed antiarrhythmic drugs, one prior radiofrequency ablation procedure, or failure of adjunctive therapies such as mechanical circulatory support or sympathetic blockade. An independent electrophysiologist provided independent verification of the treatment decision.

### 2.2 Electroanatomical Mapping and Imaging

Electroanatomical mapping data from EnSite Precision or CARTO systems were used to define arrhythmogenic substrates. Myocardial scars were identified as regions of low voltage (<0.5 mV bipolar) and validated using gadolinium-enhanced MRI or contrast-enhanced CT. These datasets were fused to optimize target delineation and ensure comprehensive coverage of the arrhythmogenic substrate.

## 2.3 Treatment Planning

Planning images were obtained using the average CT derived from 4DCT scans to account for cardiac and respiratory motion. To ensure reproducibility, patient positioning was standardized using an SBRT wing board. Photon STAR plans were created using volumetric modulated arc therapy (VMAT) with 25 Gy prescribed in a single fraction, and clinical target volumes (CTVs) were expanded by a margin (e.g., 5mm) to account for positional uncertainties, as described by Lloyd et al (18). Retrospective proton plans were developed using robust optimization in RayStation, applying the same margins from photon CTVs to proton optimization to ensure consistency. Proton range uncertainty margins were set at ±3.5%, following institutional clinical guidelines. To ensure an equitable comparison, proton plans were scaled to match the mean target volume dose of the corresponding photon plans. OAR dose constraints were established based on institutional protocols and AAPM TG-101 recommendations (19).

## 2.4 Dosimetric Comparisons

Dose-volume histograms (DVHs) were analyzed to compare dosimetric parameters between photon and proton plans. Key target metrics included $D_{98}$, $D_{95}$, and $V_{25Gy}$ to assess dose coverage and high-dose conformity for the VAS. Conformity index (CI) (20) at the 95% isodose level was calculated to assess spatial precision, and homogeneity index (HI) was used to evaluate dose uniformity within the PTV. For OARs, analyzed metrics included mean dose ($D_{mean}$), maximum dose to the hottest 0.03 cc ($D_{0.03cc}$), and volume-based metrics ($V_{5Gy}$, $V_{10Gy}$, and $V_{25Gy}$) to evaluate both low- and high-dose exposure. Specific emphasis was placed on critical thoracic structures, including the heart, lungs, esophagus, stomach, and spinal cord. Additional analyses were performed to assess doses to cardiac devices in patients with implantable devices, ensuring compliance with established safety thresholds. To contextualize potential toxicity risks, dosimetric results were further evaluated against single-fraction SBRT dose constraints outlined by Timmerman (21).

### 2.5 Statistical Analysis

Normality of dosimetric data was assessed using Shapiro-Wilk tests. Depending on the distribution, paired t-tests were used for normally distributed data, while Wilcoxon signed-rank tests were applied for non-normally distributed data. Multiple comparisons were adjusted using the Bonferroni correction method. Statistical significance was defined as $p < 0.05$. All analyses were conducted using Python (version 3.9.11) and SciPy (version 1.8.1).

## 3. Results

### 3.1 Patient Characteristics

Table 1 provides an overview of the patient cohort, including demographics, comorbidities, and treatment-specific details. The median patient age was 64.5 years (range: 48–84 years), with 26 males and 7 females. Most patients were white (22 out of 33), and almost all (90.9%) had an implantable ICD. The VAS had a median volume of 29.7 cc (range: 5.17–104.34 cc), reflecting the variability in target sizes.

### 3.2 Dosimetric Comparisons

#### 3.2.1 Target Coverage

The dosimetric analysis demonstrated that photon therapy provided slightly better target coverage, with higher D99 (21.4 ± 2.8 Gy for protons vs. 22.6 ± 2.4 Gy for photons, p = 0.022), comparable D95 coverage (24.1 ± 1.2 Gy vs. 24.7 ± 1.0 Gy, p = 0.294), and higher V25Gy (85.6 ± 10.2% vs. 93.0 ± 8.3%, p < 0.001). Full details are provided in Table 2.

#### 3.2.2 Organs at Risk

Proton therapy demonstrated dosimetric advantages for multiple OARs. For the heart, the mean dose was significantly lower with protons (3.6 ± 1.5 Gy) compared to photons (5.5 ± 2.0 Gy, p < 0.001). Protons delivered significantly lower doses to the lungs, esophagus, and spinal cord, with reductions observed in $D_{mean}$ and $D_{0.03cc}$, except for the lungs, where D0.03cc differences were not significant. For instance, the mean esophagus dose was 0.3 ± 0.6 Gy for protons vs. 1.6 ± 1.3

Gy for photons (p < 0.001). Chest wall received comparable doses with protons for certain metrics, such as $D_{0.03cc}$ (19.1 ± 7.2 Gy vs. 18.9 ± 7.6 Gy, p = 1.0). Cardiac device doses were minimal, but protons resulted in a significant reduction in $D_{0.03cc}$ (0.0 ± 0.2 vs. 0.3 ± 0.6, p < 0.001). A detailed breakdown of these metrics is available in Table 2.

When comparing OAR doses to single-fraction constraints established by Timmerman (21), no violations were observed for esophagus ($D_{5cc}$ < 14.5 Gy), spinal cord ($D_{0.25cc}$ < 10 Gy), or stomach ($D_{10cc}$ < 13 Gy). This indicates that both photon and proton STAR plans adhered to conservative normal tissue dose limits, further supporting the safety of this approach.

### 3.3 Case Studies: Patient-Specific Dosimetric Comparisons

Figures 1-3 demonstrate the dosimetry comparisons between proton and photon plans for 3 distinct patients. Figure 1 illustrates the dose distributions and DVHs for a patient with a VAS located in the superior-posterior region of the left ventricle. Both proton and photon plans achieved comparable target coverage, with VAS $D_{95}$ values of 25.51 Gy for protons and 24.85 Gy for photons. The CI was nearly identical for the two modalities (0.96 vs. 0.97), reflecting high precision in covering the target volume. Notably, proton therapy demonstrated substantial benefits in sparing adjacent OARs. Proton therapy reduced the mean heart dose by nearly half (2.2 Gy vs. 4.2 Gy). Low-dose heart exposure, represented by $V_{5Gy}$, was also minimized (11% vs. 24%; a reduction of over 50%). Additionally, the mean esophageal dose was lower with protons (2.2 Gy vs. 4.2 Gy), highlighting superior sparing of this critical structure.

Figure 2 presents dose distributions and DVHs for a patient with two VAS: one intersecting the ascending aorta and left ventricle (LV), and the other located in the anterior region of the LV. Both proton and photon plans achieved excellent target coverage, with VAS $D_{95}$ values of 23.5 Gy for protons and 24.8 Gy for photons. The CI was comparable for both modalities at 0.91 vs. 0.99, highlighting precise dose coverage of the target volumes. Despite comparable target coverage, proton therapy showed improved sparing of critical cardiac structures. The mean heart dose was lower with protons (3.3 Gy vs. 7.0 Gy for photons), accompanied by a notable reduction in low-dose heart exposure ($V_{5Gy}$: 18.7% vs. 48.2%). Additionally, proton therapy provided better sparing of the right ventricle (RV), which was adjacent to one of the substrates. The mean dose to the RV was reduced from 5.9 Gy with photons to 0.4 Gy with protons. Similarly, low-dose exposure to the lungs was greatly reduced with protons (lungs $V_{5Gy}$: 0.2% vs. 3.4% for photons).

For the third representative case (Figure 3), the VAS was located in two distinct regions: one at the junction of the left and right ventricles and the other in the inferior-posterior segment of the left ventricle, positioned near the stomach. Proton and photon plans demonstrated comparable target coverage (VAS $D_{95}$: 24.2 Gy vs. 24.81 Gy), with similar conformity index values (CI: 0.96 vs. 0.98). However, proton therapy provided better sparing of nearby OARs. Proton therapy lowered the mean dose to the stomach (0.1 Gy vs. 1.00 Gy). The spinal cord received considerably lower doses with protons ($D_{mean}$: 0.3 cGy vs. 0.6 Gy, $D_{0.03cc}$: 3.4 cGy vs. 6.0 Gy).

**Table 1**: Patient characteristics for the cohort undergoing stereotactic arrhythmia radioablation (STAR) using protons and photons. Values are expressed as medians with ranges where applicable. ICD percentages represent the proportion of patients with implanted devices.

| Characteristic | Value |
|---|---|
| Number of Patients | 33 |
| Age | 64 (48, 84) |
| Gender | 26 Male/ 7 Female |
| Race | 22 White/11 Others |
| Median VAS volume (cc) | 68.25 (20.47, 239.36) |
| ICD (%) | 90.91 |
| **VAS:** Ventricular Arrhythmia Substrate, **ICD:** Implantable Cardioverter-Defibrillator | |

**Table 2**: Dosimetric comparisons between proton and photon treatment plans for various target and OAR metrics. Values are expressed as mean ± standard deviation (SD), with significant results highlighted.

| Structure | DVH Metric | Proton (Mean ± SD) | Photon (Mean ± SD) | p-value |
|---|---|---|---|---|
| VAS | $D_{99}$ (Gy) | 21.4 ± 2.8 | 22.6 ± 2.4 | 0.022 |
| | $D_{95}$ (Gy) | 24.1 ± 1.2 | 24.7 ± 1.0 | 0.294 |
| | $V_{25Gy}$ (%) | 85.6 ± 10.2 | 93.0 ± 8.3 | <0.001 |
| Whole Heart with VAS | $D_{mean}$ (Gy) | **3.6 ± 1.5** | **5.5 ± 2.0** | **<0.001** |
| | $D_{0.03cc}$ (Gy) | 31.9 ± 1.8 | 31.5 ± 1.4 | 1.000 |
| Normal Heart without VAS | $D_{mean}$ (Gy) | **2.4 ± 1.2** | **4.4 ± 1.8** | **<0.001** |
| | $V_{5Gy}$ (%) | **15.7 ± 8.5** | **32.5 ± 17.1** | **<0.001** |
| Lungs | $D_{mean}$ (Gy) | **0.7 ± 0.6** | **1.2 ± 0.7** | **<0.001** |
| | $V_{5Gy}$ (%) | 5.1 ± 4.7 | 6.3 ± 5.1 | 0.153 |
| Right Lung | $D_{mean}$ (Gy) | **0.0 ± 0.0** | **0.6 ± 0.3** | **<0.001** |
| | $V_{5Gy}$ (%) | 0.0 ± 0.2 | 0.1 ± 0.3 | 1.000 |
| Chest Wall | $D_{0.03cc}$ (Gy) | 19.1 ± 7.2 | 18.9 ± 7.6 | 1.000 |
| | $V_{25Gy}$ (%) | 0.6 ± 1.3 | 0.5 ± 1.1 | 1.000 |
| Esophagus | $D_{mean}$ (Gy) | **0.3 ± 0.6** | **1.6 ± 1.3** | **<0.001** |
| | $D_{0.03cc}$ (Gy) | **3.2 ± 4.9** | **7.0 ± 3.4** | **<0.001** |
| Spinal Cord | $D_{0.03cc}$ (Gy) | **0.2 ± 1.3** | **3.2 ± 1.5** | **<0.001** |

| Stomach | $D_{mean}$ (Gy) | 0.2 ± 0.3 | 0.7 ± 0.9 | <0.001 |
| Cardiac Device | $D_{0.03cc}$ (Gy) | 0.0 ± 0.2 | 0.3 ± 0.6 | <0.001 |
| $D_{XX}$: Dose received by XX% of the volume, $V_{YGy}$: Volume receiving at least Y Gy, $D_{mean}$: Mean dose, $D_{0.03cc}$: Hottest 0.03 cc dose, CI: Conformity Index, HI: Homogeneity Index ||||||

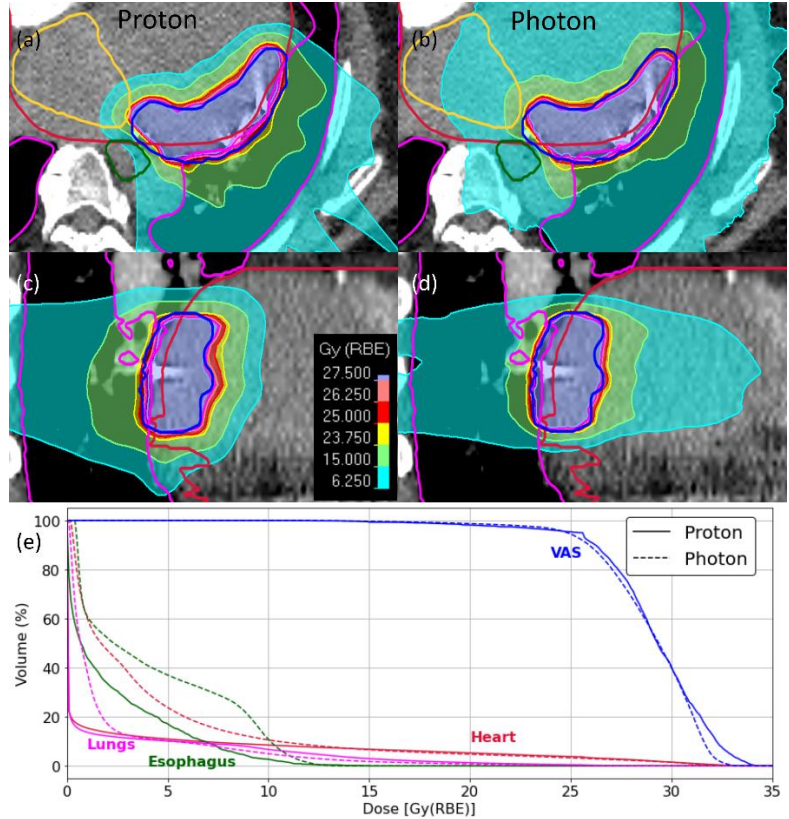

**Figure 1:** Dose distributions and dose-volume histograms (DVHs) comparing proton and photon plans for a representative patient treated with STAR. **(a, c)** Axial and sagittal views for the proton plan. **(b, d)** Axial and sagittal views for the photon plan. Contours for the ventricular arrhythmogenic substrate (VAS, blue), heart (crimson), esophagus (dark green), lungs (magenta), and left atrium (yellow) are overlaid. Thicker structural boundaries were used to distinguish structures from isodose lines where colors overlap. **(e)** DVHs demonstrate comparable VAS coverage and superior sparing of the esophagus, left atrium, and normal heart with protons.

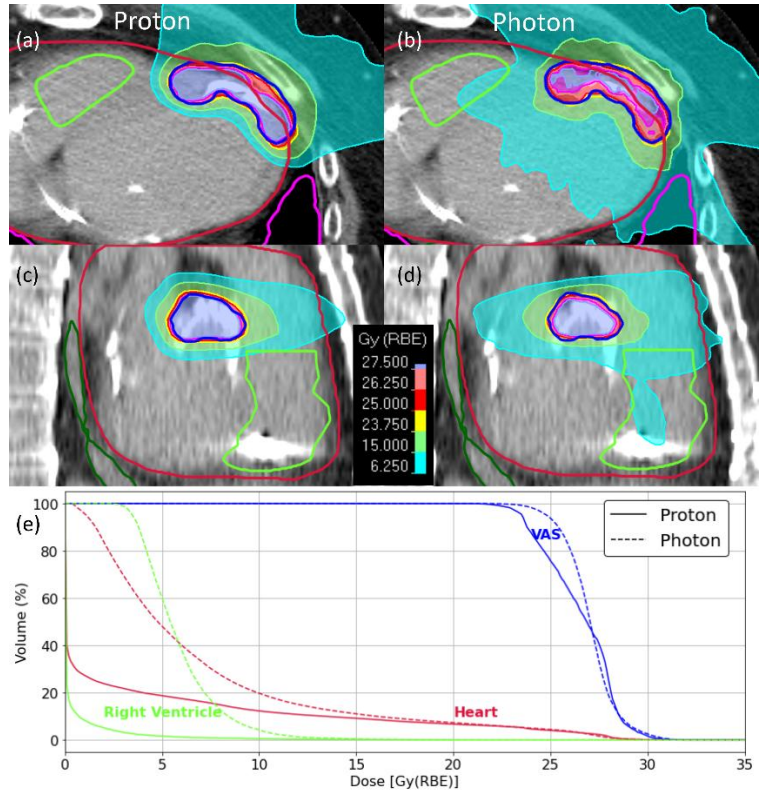

**Figure 2:** Dose distributions and dose-volume histograms (DVHs) comparing proton and photon plans for a patient with the ventricular arrhythmogenic substrate (VAS) located near the right ventricle. **(a, c)** Axial and sagittal views for the proton plan. **(b, d)** Axial and sagittal views for the photon plan. Contours for the VAS (blue), heart (crimson), and right ventricle (bright green) are overlaid. Thicker contour boundaries were used to distinguish structures from isodose lines. **(e)** DVHs illustrate superior sparing of the right ventricle and heart with proton therapy, while maintaining comparable VAS coverage.

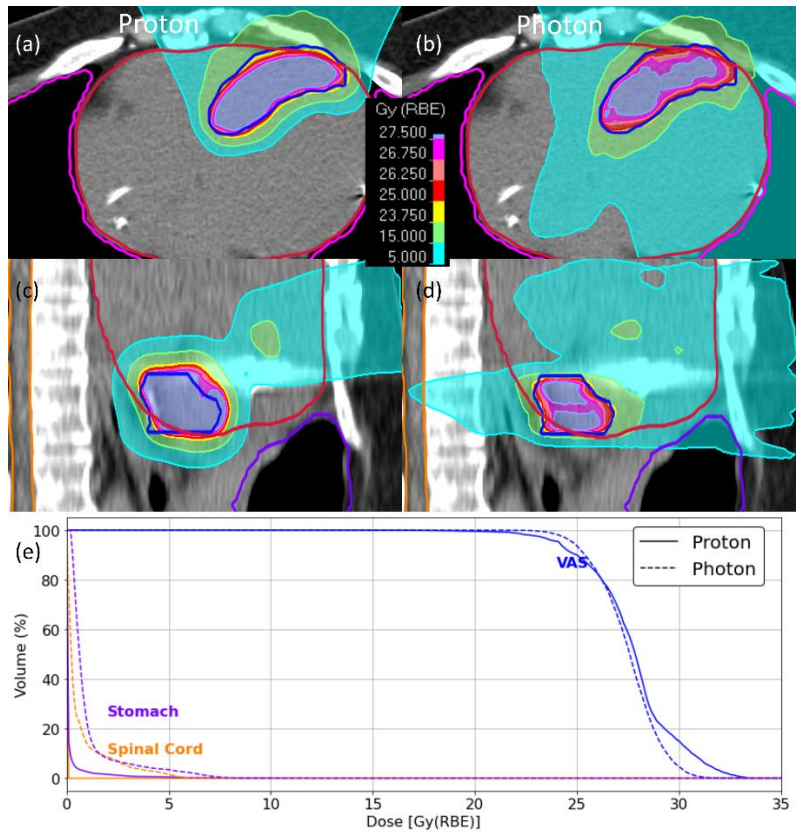

**Figure 3:** Dose distributions and dose-volume histograms (DVHs) for a patient with the ventricular arrhythmogenic substrate (VAS) located near the left ventricle. **(a, c)** Axial and sagittal views for the proton plan. **(b, d)** Axial and sagittal views for the photon plan. Contours for the VAS (blue), heart (crimson), stomach (purple), and spinal cord (orange) are overlaid, with thicker contour boundaries distinguishing structures from isodose lines. **(e)** DVHs demonstrate comparable VAS coverage with superior sparing of the stomach and spinal cord using proton therapy compared to photons.

## 4. Discussion

Our study highlights the potential of proton therapy to enhance dosimetric outcomes for STAR in patients with VT. Leveraging the unique physical properties of protons, including the Bragg peak, enabled superior sparing of critical OARs without compromising VAS target coverage. These findings underscore the role of proton STAR in optimizing treatment for high-risk patients, particularly those with complex cardiac anatomy or significant comorbidities.

Proton therapy demonstrated comparable target coverage to photons, with similar CI and dose coverage metrics, including $D_{95}$ and $D_{98}$. However, significant reductions in mean and high-dose exposure to critical OARs were observed with protons, including the heart, lungs, esophagus, spinal cord, and stomach. For instance, mean esophageal dose was reduced by more than 50% in most cases, while low-dose exposure to the lungs and heart (e.g., $V_{5Gy}$) was consistently minimized. These results align with thoracic oncology studies, such as RTOG 1308, which demonstrated superior OAR sparing with protons compared to photons in lung cancer patients (22). The precision and conformity of proton therapy make it particularly well-suited for STAR, where critical structures often lie in close proximity to the target.

The ability of proton therapy to spare critical thoracic structures has profound implications for minimizing acute and late toxicities in VT patients. Reducing esophageal dose can decrease the risk of radiation-induced esophagitis, while lower lung doses are likely to mitigate pneumonitis, particularly for patients with pre-existing pulmonary conditions or reduced baseline lung function. Additionally, reducing dose to cardiac substructures, such as the right ventricle and coronary arteries, may lower the risk of long-term cardiac dysfunction and arrhythmias. This is particularly relevant in patients with compromised cardiac health or prior radiation exposure, where minimizing cumulative dose to the heart is critical. Our findings are consistent with prior studies in thoracic oncology (23–26) which demonstrated significant reductions in cardiac and pulmonary doses with proton therapy compared to photon-based SBRT or IMRT. Specifically, studies have reported reductions in mean heart dose, lung $V_{5Gy}$, and esophageal dose with protons, which align with our observations of superior OAR sparing in STAR for VT. These dose reductions are clinically meaningful in mitigating risks of esophagitis and pneumonitis, which are particularly critical for VT patients who are already high-risk due to their compromised cardiac function and underlying structural heart disease. While STAR is designed to modulate arrhythmogenic substrates, care must be taken to minimize inadvertent dose to adjacent cardiac structures, as emerging data suggest that high-dose radiation to certain cardiac substructures could potentially exacerbate arrhythmias in vulnerable patients (27–30).

Beyond dosimetric advantages, proton therapy has also been shown to reduce systemic toxicities, as highlighted by Cortiula *et al*, who reported lower rates of hematological toxicities and improved treatment tolerance in patients undergoing concurrent chemoradiotherapy for stage III NSCLC (31). While not directly applicable to VT, this finding underscores the potential systemic benefits of proton therapy, which could hold relevance for VT patients requiring ongoing medical management or implantable cardiac devices.

Patient selection remains a critical consideration for maximizing the benefits of proton therapy. Studies like Teoh *et al* demonstrated that patients with pre-existing cardiac disease or targets near critical structures, such as the heart or esophagus, derive the greatest benefit from protons (32). Similarly, Zientara *et al* reviewed global practices for proton therapy selection, highlighting model-based approaches, such as normal tissue complication probability (NTCP) modeling and dosimetry comparisons, as key tools for optimizing treatment in anatomically challenging cases (33). These methodologies could be extended to STAR to ensure optimal patient outcomes, particularly in scenarios where photon therapy may result in excessive OAR doses.

Pediatric patients with VT represent a unique and challenging population due to their increased susceptibility to late toxicities and longer life expectancy. A case report by Lee *et al* (34) demonstrated the feasibility of STAR in an 11-year-old with refractory VT and severe dilated cardiomyopathy. Despite multiple failed radiofrequency catheter ablation (RFCA) attempts and intolerance to antiarrhythmic drugs, photon-based STAR successfully reduced the VT burden without significant adverse effects. Proton therapy could further improve outcomes in such cases by reducing dose to surrounding tissues, as highlighted in studies by Lee *et al* (35), Chaikh *et al* (36) and Bates *et al* (37, 38). These studies underscore the importance of minimizing radiation dose to critical structures, including cardiac substructures and lungs, to reduce the risk of secondary malignancies, growth disturbances, and late cardiac disease.

Range uncertainties remain a challenge in proton therapy, particularly in anatomically complex cases where critical structures lie near the distal edge of the proton beam. Studies by Seco *et al*. have demonstrated that while protons offer significant OAR sparing, range uncertainties can result in slightly larger high-dose regions compared to photons when fewer beams are used (39, 40). Future studies should explore advanced techniques, including pencil beam scanning and proton arc therapy, to mitigate these uncertainties and improve dose distributions. Another important consideration in proton therapy is the uncertainty surrounding the relative biological effectiveness (RBE). While a fixed RBE of 1.1 is commonly used in clinical practice, studies like Underwood *et al* suggest that RBE values may be higher at the distal edge of the proton spread-out Bragg peak, particularly for late-responding tissues such as the lungs (41). This variability could have significant implications for STAR, where critical structures often lie near the distal edge of the proton beam. Paganetti further emphasizes that RBE is influenced by several physical and biological factors, including dose, linear energy transfer (LET), and tissue type, underscoring the need for more personalized RBE modeling in treatment planning (42, 43). Incorporating LET-based planning and RBE-adaptive approaches could optimize dose delivery in STAR and minimize unexpected toxicities. Inaccuracies in substrate delineation can significantly impact the efficacy of STAR, as demonstrated by Haskova *et al*, who reported cases of repeated SBRT due to imprecise initial targeting (44). These observations emphasize the need for advanced substrate identification strategies, such as co-registration of electroanatomical maps with CT imaging, to ensure precise targeting. Incorporating machine learning and auto-segmentation algorithms could further enhance reproducibility and accuracy in treatment planning, particularly for complex or intramural substrates.

Emerging technologies, such as MR-guided STAR, offer additional avenues for improving treatment precision. Studies by Bianchi *et al* (45) and Akdag *et al* (46) have highlighted the feasibility of integrating real-time motion tracking and adaptive planning into STAR workflows, even in patients with implantable devices. These advances could enhance target coverage and OAR sparing by accounting for dynamic changes in cardiac and respiratory motion during treatment.

One notable limitation of proton therapy is the potential impact of secondary neutrons on ICDs and pacemakers. Studies by Oshiro *et al* (47), Hashimoto *et al* (48) and Makkar *et al* (49) reported transient malfunctions, such as power-on resets, in ICDs exposed to neutron scatter during proton therapy, with an approximate 1 in 50 Gy probability of occurrence. While these events were infrequent and did not result in permanent damage, proactive strategies, including real-time monitoring and adherence to AAPM TG-203 guidelines, are essential to ensuring patient safety during proton STAR (50). Further work is needed to quantify and mitigate neutron exposure to ICDs, with Monte Carlo-based modeling offering a potential avenue to improve dose estimation and refine treatment planning. This would allow for a more comprehensive risk assessment and the development of strategies to minimize secondary neutron effects in future proton STAR applications.

Ongoing clinical trials, such as the Mayo Clinic study (NCT04392193) and the TOVEL study (NCT06769451), will provide valuable insights into the feasibility and efficacy of proton STAR for VT. These studies complement preclinical research that has demonstrated the potential of protons to deliver conformal and homogeneous doses to VAS while sparing surrounding OARs. As data from these trials become available, they will likely shape future clinical practice and support the broader adoption of proton STAR.

Future research should also prioritize long-term follow-up to evaluate the impact of reduced OAR doses on survival and toxicity. Multi-institutional collaborations and larger cohort studies are essential to validate our findings and assess their generalizability. Additionally, incorporating cardiac substructure contouring and advanced RBE models into treatment planning could further refine proton STAR for both pediatric and adult patients.

This study is limited by its retrospective design and small sample size, which may introduce selection bias. Additionally, our analysis focused on dose metrics for larger cardiac structures without incorporating detailed substructure contouring, which is an area for future research. Prospective studies with larger cohorts and multi-institutional collaboration are needed to validate our findings and assess their impact on clinical outcomes. Furthermore, long-term follow-up is essential to evaluate the effects of reduced OAR doses on late toxicities and survival.

## 5. Conclusion

Our study demonstrates the dosimetric advantages of proton therapy in sparing OARs while maintaining optimal target coverage. These results support the adoption of safer and more effective treatment strategies for VT patients and establish a dosimetric foundation for utilizing proton STAR in clinical practice. With superior OAR sparing and the potential for dose escalation, proton therapy is a compelling option for managing complex cardiac arrhythmias, particularly in pediatric and high-risk adult populations. Ongoing advancements in imaging, planning, and delivery technologies, combined with prospective clinical validation, will further define the role of proton STAR in this evolving field.

**Acknowledgments**

This research is supported in part by the National Institutes of Health under Award Number R01CA272991, R56EB033332, R01EB032680, R01DE033512, R37CA272755 and U54CA274513.